\documentclass[%
reprint,
 amsmath,amssymb,
 aps,
]{revtex4-2}
\usepackage{siunitx}
\usepackage{graphicx}% Include figure files
\usepackage{dcolumn}% Align table columns on decimal point
\newcommand{\RNum}[1]{\uppercase\expandafter{\romannumeral #1\relax}}
\usepackage{wrapfig}
\usepackage{float}
\usepackage{bm}% bold math
\usepackage[dvipsnames]{xcolor}
\usepackage[linktoc=page, colorlinks=true, linkcolor=MidnightBlue, citecolor=MidnightBlue, urlcolor=OliveGreen]{hyperref}
\begin{document}

\preprint{APS/123-QED}
\title{Electronic structure of InAs and InSb surfaces: density functional theory and angle-resolved photoemission spectroscopy}

%\RomanL{another title suggestion: Electronic structure of InAs and InSb surfaces: density functional theory and angle-resolved photoemission spectroscopy perspective}

\author{Shuyang Yang}
\affiliation{Department of Materials Science and Engineering, Carnegie Mellon University, Pittsburgh, PA 15213, USA
}%
\author{Niels B. M. Schröter}
\affiliation{Paul Scherrer Institut, Swiss Light Source, CH-5232 Villigen PSI, Switzerland}
\author{Sergej Schuwalow}
\affiliation{Center for Quantum Devices, Niels Bohr Institute, University of Copenhagen and Microsoft Quantum Materials Lab Copenhagen, Lyngby, Denmark}
\author{Mohana Rajpalk}
\affiliation{Microsoft Quantum Materials Lab Copenhagen, Lyngby, Denmark}

\author{Keita Ohtani}
\affiliation{Microsoft Quantum Materials Lab Copenhagen, Lyngby, Denmark}

\author{Peter Krogstrup}
\affiliation{Microsoft Quantum Materials Lab Copenhagen, Lyngby, Denmark}

\author{Georg W. Winkler}
\affiliation{Station Q, Microsoft Corporation, Santa Barbara, California 93106-6105, USA}
\affiliation{Microsoft Quantum, One Microsoft Way Redmond, WA 98052, USA}
\author{Jan Gukelberger}
\affiliation{Microsoft Quantum, One Microsoft Way Redmond, WA 98052, USA}
\author{Dominik Gresch}
\affiliation{Station Q, Microsoft Corporation, Santa Barbara, California 93106-6105, USA}
\author{Gabriel Aeppli}
\affiliation{Physics Department, ETH CH-8093, Zurich, Switzerland}
\affiliation{Institut de Physique, EPFL CH-1015 , Lausanne, Switzerland}
\affiliation{Paul Scherrer Institut, Swiss Light Source, CH-5232 Villigen PSI, Switzerland}
\author{Roman M. Lutchyn}
\affiliation{Station Q, Microsoft Corporation, Santa Barbara, California 93106-6105, USA}
\affiliation{Quantum Science Center, Oak Ridge, TN 37830 USA}
\author{Vladimir N. Strocov}
\affiliation{Paul Scherrer Institut, Swiss Light Source, CH-5232 Villigen PSI, Switzerland}
\author{Noa Marom}
\affiliation{Department of Materials Science and Engineering, Carnegie Mellon University, Pittsburgh, PA 15213, USA
}%
\affiliation{Department of Physics, Carnegie Mellon University, Pittsburgh, PA 15213, USA
}%
\affiliation{Department of Chemistry, Carnegie Mellon University, Pittsburgh, PA 15213, USA
}%
\date{\today}% It is always \today, today,

\begin{abstract}
The electronic structure of surfaces plays a key role in the properties of quantum devices. However, surfaces are also the most challenging to simulate and engineer. Here, we 
study the electronic structure of InAs(001), InAs(111), and InSb(110) surfaces using a combination of density functional theory (DFT) and angle-resolved photoemission spectroscopy (ARPES). We were able to perform large-scale first principles simulations and capture effects of different surface reconstructions by using DFT calculations with a machine-learned Hubbard U correction [npj Comput. Mater. 6, 180 (2020)]. To facilitate direct comparison with ARPES results, we implemented a "bulk unfolding" scheme by projecting the calculated band structure of a supercell surface slab model onto the bulk primitive cell. 
For all three surfaces, we find a good agreement between DFT calculations and ARPES. For InAs(001), the simulations clarify the effect of the surface reconstruction. Different reconstructions are found to produce distinctive surface states. For InAs(111) and InSb(110), the simulations help elucidate the effect of oxidation. Owing to larger charge transfer from As to O than from Sb to O, oxidation of InAs(111) leads to significant band bending and produces an electron pocket, whereas oxidation of InSb(110) does not. Our combined theoretical and experimental results may inform the design of quantum devices based on InAs and InSb semiconductors, e.g., topological qubits utilizing the Majorana zero modes. 
 
\end{abstract}
\maketitle

%\tableofcontents
\section{\label{sec:level1}INTRODUCTION}

The narrow-gap III-V semiconductors InAs and InSb (InX) have attractive material parameters, including small effective mass, large Lande $g$-factor, and large spin-orbit coupling. This unique combination of material parameters has generated considerable interest in InX semiconductors for many technological applications. Thanks to the small electron effective mass and long mean free path, InX materials have high electron mobility, which is important for applications that require high speed and low noise at very low bias voltage~\cite{boos1998alsb,mcconville1994direct, ashley1991ambient}.  Furthermore, InX materials are relevant to infrared detectors thanks to their direct gap and high sensitivity in the atmospheric window between 3-5 $\mu$m~\cite{cohen2017short,ueno2013insb}. InX materials are also promising for spintronics~\cite{Awschalom2009,Premasiri_2019,MartinezPRA2020} because of their strong spin-orbit interaction and gate-tunable Rashba coefficients~\cite{10.1021/nl301325h, leontiadou2011experimental,shojaei2016demonstration}.

Recently, InX materials have attracted considerable attention as promising platforms for realizing Majorana zero modes~\cite{Sau2010, Alicea10, Lutchyn2010, Oreg2010,  Alicea2012, Beenakker2013, Flensberg_review, DasSarma2015,Lutchyn17,Aguado17}. The main goal of these proposals~\cite{Sau2010, Alicea10, Lutchyn2010, Oreg2010} is to engineer topological p-wave superconductivity at the interface of a conventional semiconductor and an s-wave superconductor. Exceptional control of interface properties is necessary to realize topological superconducting phases and manipulate Majorana zero modes, key requirements in topological quantum computation proposals~\cite{Nayak2008,DasSarma2015, Lutchyn17}. The semiconductor in such a hybrid semiconductor-superconductor Majorana device is required to have a large $g$-factor, strong Rashba spin-orbit coupling and significant proximity-induced superconducting gap. Proximity-induced superconductivity has been studied recently in nanowires, including NbTiN/InSb~\cite{Mourik2012, Zhang2016, Guel2017, su2020erasing,chen2019ubiquitous,chen2017experimental,zuo2017supercurrent,su2018mirage}, Al/InAs~\cite{Krogstrup15, suominen2017zero,nichele2017scaling,matsuo2020evaluation,Krogstrup15,vaitiekenas2018selective, menard2020conductance, Vaitiekenas2020}, Al/InSb~\cite{gazibegovic2017epitaxy, anselmetti2019end,de2018electric,shen2018parity,het2020plane}, Pb/InAs~\cite{kanne2020epitaxial}, and Sn/InSb~\cite{pendharkar2019paritypreserving}.
High-quality, uniform and transparent superconductor-semiconductor interfaces are required to optimize topological gaps in semiconductor-superconductor heterostructures~\cite{Stanescu2011, Antipov2018, Winkler2019}.

The functionality and performance of InX hybrid interfaces in the aforementioned applications depend critically on the interface quality. Changes in the structure of the interface at the atomic scale affect the electronic properties. Indeed, the interface configuration may give rise to (desirable or undesirable) interface states, alter the band bending and band alignment, or affect the magnitude of the proximity-induced gap and spin-orbit coupling \cite{doi:10.1063/1.368355,doi:10.1021/acsomega.9b03362, PhysRevB.100.085128}. A deeper understanding of the relation between the structure and electronic properties of InX hybrid interfaces could advance the synthesis of precisely controlled interfaces with tunable properties. 

InX hybrid interfaces are often grown by molecular beam epitaxy (MBE) with InX serving as the substrate \cite{doi:10.1002/1521-3951(200201)229:1<19::AID-PSSB19>3.0.CO;2-J,Krogstrup15,kanne2020epitaxial,kanne2020epitaxial,gazibegovic2017epitaxy,chen2017experimental}. The nanowire-based Majorana devices typically have multiple interfaces, including InX-superconductor and InX-oxide interfaces. Therefore, elucidating the structure and electronic properties of InAs and InSb surfaces is an important first step towards advancing the understanding of epitaxially grown InX hybrid interfaces.
InAs and InSb exhibit a number of surface reconstructions, which have been observed with atomic resolution using scanning tunneling microscopy and low-energy electron diffraction (LEED) \cite{ratsch2000surface, bracker2000surface, liu1994surface, mcconville1994surface, schmidt2002iii}. Understanding the resulting surface states and Fermi-level pinning is important for engineering appropriate interface Hamiltonian and realizing topological superconductivity hosting Majoranas. Surface reconstructions with less pinning and no problematic surface states close to the band edges are preferred \cite{Winkler2019}. It may be challenging to identify the InX surface reconstructions present in experiments because some samples may contain highly reconstructed surfaces with large unit cells, disorder, or domains with related but different reconstructions \cite{goryl2010structure,goryl2011structure}.  Band bending and surface states have been observed by angle-resolved photoemission spectroscopy (ARPES)\cite{tomaszewska2015surface, schuwalow2019band}, however it is a non-trivial task to assign spectroscopic signatures to specific structural features. For instance, without comparing to calculations, it is difficult to associate surface state dispersion with the corresponding surface reconstruction~\cite{PhysRevLett.47.443}. Moreover, a number of experiments have investigated the effect of the adsorption of gases on the surfaces of InAs and observed the formation of an accumulation layer, for instance for $N_2$ on InAs \cite{1993_Gemmeren_ApplSurfSci} or $O_2$ on InAs \cite{1986_Baier_SolidStateCommun, 2003_Leandersson_ApplSurfSci}. However, the microscopic origin of the charge accumulation has not been understood. Furthermore, there has been no systematic comparison of oxygen adsorption between oxidized InAs and InSb surfaces, %which is particularly important for the understanding of 
as it relates to the electronic structure of proximitized nanowire devices, which typically have some facets covered by a native oxide.

%In this paper we use DFT+U(BO) to investigate the electronic structure of InAs(001), InAs(111) and InSb(110) surfaces. 
First principles simulations based on density functional theory (DFT) can help interpret experiments and resolve the effects of surface reconstructions and defects \cite{PhysRevB.80.201309,DEOLIVEIRA20101319, doi:10.1021/jacs.7b01081,doi:10.1021/acsnano.9b02996,C7EE03482B,YANG2018184}. DFT studies of InAs and InSb surfaces and interfaces have been limited because local and semi-local exchange-correlation functionals severely underestimate the band gap to the point that it reduces to zero \cite{10.1063/1.2404663}. Refs. \cite{ratsch2000surface, yeu2017surface} focused on the InAs(001) surface reconstruction equilibrium phase diagram using the local density approximation (LDA). InAs(110) and InAs(001) surfaces have been studied using LDA with empirically modified pseudo-potentials in Refs. \cite{10.1063/1.3518061, miwa2000structure}.
  Hybrid functionals, such as the the Heyd-Scuseria-Ernzerhof (HSE) functional \cite{Heyd2003,doi:10.1063/1.2204597}, which contain a fraction of the exact exchange, yield band gaps in good agreement with experimental values for InAs and InSb \cite{Kresse2009, kosmider2013electronic}. However, it is not feasible to use hybrid functionals for simulations of large surface and interface models due to their high computational cost. One alternative to hybrid functionals is the modified Becke-Johnson (mBJ) semilocal exchange \cite{PhysRevLett.102.226401}, which has a local approximation to an atomic exact-exchange potential. The mBJ functional can produce band gap values similar to hybrid functionals for InAs and InSb at a lower computational cost \cite{wang2013electronic}. However, the mBJ functional can  only be used for fully periodic systems. For surface and interface calculations with a vacuum region mBJ calculations tend to diverge because in the vacuum, where the electron density and kinetic energy density are close to zero, the functional becomes unstable \cite{eremeev2019surface}. 

Recently, we have introduced a new method of DFT with a machine-learned Hubbard U correction, which can provide a solution for accurate and efficient simulations of InAs and InSb~\cite{yu2020machine}. Within the Dudarev formulation of DFT+U \cite{PhysRevB.57.1505} the effective Hubbard U is defined as $U_{eff}=U-J$ with $U$ and $J$ representing the on-site Coulomb repulsion and the exchange interaction, respectively. For a given material, the $U_{eff}$ parameters of each element are machine-learned using Bayesian optimization (BO). The BO algorithm finds the optimal $U_{eff}$ values that maximize an objective function formulated to reproduce as closely as possible the band gap values and the qualitative features of the band structure obtained with a hybrid functional.  %Unlike linear response method which is widely used to determine $U_{eff}$ values,
The DFT+U(BO) method allows for negative $U_{eff}$ values. Negative $U_{eff}$ values are theoretically permissible when the exchange term, $J$, is larger than the on-site Coulomb repulsion, $U$ \cite{RevModPhys.62.113,hase2007madelung,nakamura2009first,persson2006improved, cococcioni2012lda+}. We have found that negative $U_{eff}$ values are necessary to produce band gaps for narrow-gap semiconductors, such as InAs and InSb. Because the reference hybrid functional calculation is performed only once for the bulk material to determine the optimal $U_{eff}$ values, the computational cost of DFT+U(BO) calculations for surfaces is comparable to semi-local DFT. Moreover, unlike the mBJ functional, DFT+U(BO) can be used for surface and interface models with a vacuum region. 

Supercells are often used in DFT simulations to study the effect of surface reconstructions, disorder, and defects within a periodic picture. The complexity of a supercell band structure makes it difficult to compare with ARPES experiments. Band unfolding \cite{popescu2012extracting, 2010_Ku_PRL} is a useful method for projecting the band structure of a supercell onto the corresponding primitive cell with appropriate spectral weights. Previously, band unfolding has been applied to surface slab band structures only in the $xy$ plane, i.e., a supercell of X$\times$Y$\times$Z was unfolded onto the corresponding 1$\times$1$\times$Z slab. However, when the band structure of a slab with a large number of layers is unfolded only in the $xy$ plane, the number of bands may still be too large to compare with ARPES. Moreover, bands corresponding to non-zero values of $k_{z}$ in the primitive cell may be present in the unfolded slab band structure, creating artifacts that do not appear in ARPES. To address these issues, we have developed a "bulk unfolding" scheme, where the bands of a surface slab are unfolded onto the corresponding bulk primitive cell. This facilitate the comparison of computed band structures to ARPES experiments.

The bulk unfolding scheme is used to compare the calculated slab band structures to ARPES measurements. The results of bulk-unfolded DFT+U(BO) simulations are in good agreement with ARPES. Moreover, the simulations help interpret features observed in ARPES. For InAs(001), the effect of surface reconstruction is investigated by comparing the band structures produced by the $\beta2(2\times4)$, $\alpha2(2\times4)$, $\zeta(4\times2)$, and $c(4\times4)$ reconstructions. The results of DFT+U(BO) simulations support the presence of $\beta2(2\times4)$ and $\zeta(4\times2)$ coexisting domains. For InAs(111) and InSb(110), we investigate the effect of oxidation. In agreement with ARPES, DFT+U(BO) simulations show that the binding of oxygen to As on the InAs(111) increases the band bending and produces an electron pocket. In contrast, the binding of oxygen to Sb on InSb(110) does not significantly shift the valence band maximum and no electron pocket is observed. This is explained by larger charge transfer from As to O than from Sb to O.   

The paper is organized as follows. In Sec. II, we describe our methods, including the bulk unfolding scheme, DFT settings, and slab construction. In Sec. III experimental and theoretical results are presented for InAs(001), InAs(111), and InSb(110).  

\section{\label{sec:level1}METHODS}

\subsection{Band unfolding}

\begin{figure*}
\includegraphics[scale = 0.113]{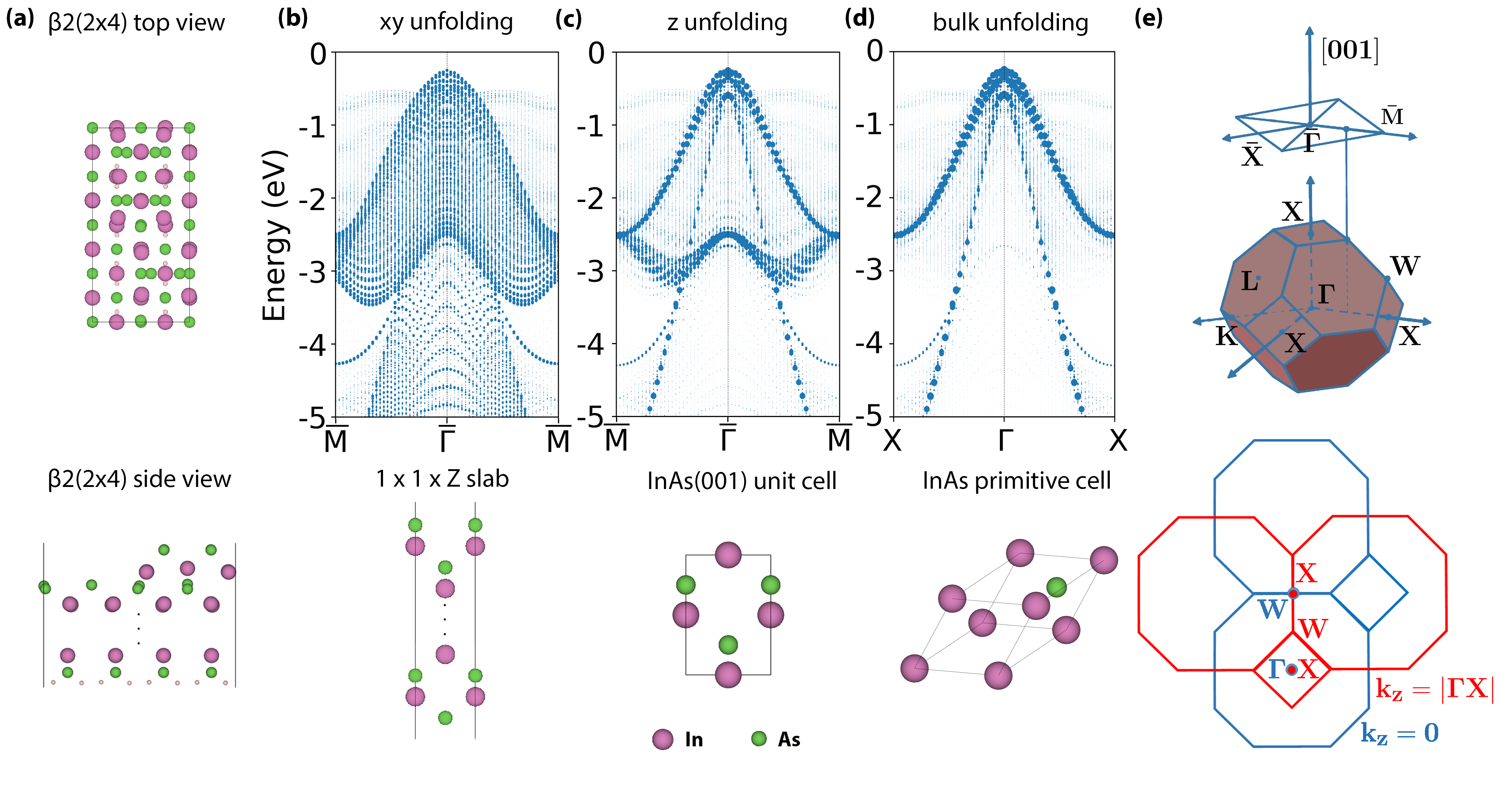}% Here is how to import EPS art
\caption{\label{fig:convert}(a) Top view and side view of the $\beta2(2\times4)$ reconstruction of InAs(001). InAs(001) $\beta2(2\times4)$ band structures with (b) $xy$-unfolding, (c) $z$-unfolding, and (d) bulk unfolding. (e) The surface Brillouin zone for InAs(001) $\beta2(2\times4)$. Red lines correspond to $k_{z} = v|\Gamma X|$ and blue lines correspond to $k_{z}=0$.}
\end{figure*}

In the following, we describe the bulk unfolding scheme. As an example, we use the InAs(001) surface with a $\beta2(2\times4)$ reconstruction, illustrated in Fig. \ref{fig:convert}a. The slab model comprises a 2$\times$4 supercell in the $xy$ plane with 20 layers and a vacuum region of 40 \AA {} in the $z$ direction. 
Band unfolding is based on the relation between the basis vectors of the supercell and the primitive cell \cite{popescu2012extracting}:
\begin{equation}
    \textbf{A} = \textbf{M} \cdot \textbf{a}
\end{equation}
where $\textbf{A}$ and $\textbf{a}$ are matrices with the supercell and primitive cell basis vectors as rows, respectively, and $\textbf{M}$ is the integer transformation matrix between the two. A similar relation holds in reciprocal space:
\begin{equation}
    \textbf{B} = \textbf{M}^{-1} \cdot \textbf{b}
\end{equation}
where $\textbf{B}$ and $\textbf{b}$ are basis vector matrices in reciprocal space corresponding to the supercell and primitive cell, respectively.
Given a wave vector $\vec{k}$  in the primitive cell Brillouin zone, there is exactly one wave vector $\vec{K}$ in the smaller supercell Brillouin zone such that the two vectors are related by a reciprocal lattice vector $\vec{G}$ of the supercell:
\begin{equation}
    \vec{k} = \vec{K} + \vec{G}
\end{equation}

The unfolded band structure is formed by weighted dots with the weights given by the spectral function of the primitive cell, $A(\vec{k}, \epsilon)$:
\begin{equation}
    A(\vec{k}, \epsilon)=\sum_{m} P_{\vec{K} m}(\vec{k}) \delta\left(\epsilon_{\vec{K} m}-\epsilon\right)
\end{equation}
where $\epsilon$ is the energy, $m$ is the band index of the supercell band structure. $P_{\vec{K} m}(\vec{k})$ is the projection of the supercell state $| \vec{K} m\rangle$ on all the primitive cell states $| \vec{k} n\rangle$ with a given $\vec{k}$ and $n$ is the primitive cell band index:
\begin{equation}
   P_{\vec{K} m}(\vec{k})=\sum_{n}|\langle\vec{K} m | \vec{k} n\rangle|^{2}
\end{equation}
 
%$A(\vec{k}, )$ is the weight at energy $\epislon$ with reciprocal space vector $\vec{k}$ and 
Because the supercell comprises several primitive cells, the supercell states may contain contributions from several primitive cell states. $P_{\vec{K} m}$ provides information about how much of the character of $|\vec{k} n\rangle$ is preserved in $|\vec{K} m\rangle$. 

Fig. \ref{fig:convert}b shows the result of the standard $xy$ unfolding scheme, obtained the VaspBandUnfolding code, available at: \url{https://github.com/QijingZheng/VaspBandUnfolding}. Within this scheme, the band structure is unfolded onto a 1$\times$1 slab in the $xy$ plane with the original dimension in the $z$ direction, illustrated under the band structure. The $xy$-unfolded band structure is still quite complex with a large number of bands due to the large supercell size in the $z$ direction. The supercell band structure may be further unfolded onto a periodic bulk unit cell with the same orientation as the surface supercell. Most band unfolding codes can, in principle, perform such "$z$-unfolding". However, the slab must be constructed such that its length in the $z$-direction is an integer multiple of the $z$-dimension of the periodic unit cell with the same orientation in order to preserve the relation between the supercell and primitive cell as described by equations (1)-(3). The $z$-unfolded band structure, shown in Fig. \ref{fig:convert}c, has fewer bands and appears more similar to the bulk band structure shown in Fig. \ref{fig:inas001hse}. However, some extra bands are present around -3 eV. 
The extra bands appear because when the primitive Brillouin zone of InAs bulk is projected onto the (001) periodic slab \cite{ibach2006physics}, the bands along $XW$ at $k_{z}$ = $|\Gamma X|$ overlap with $X\Gamma$ at $k_{z}$ = 0, as shown in Fig. \ref{fig:convert}e. Thus, the $z$-unfolding scheme eliminates the effect of the slab thickness but cannot decompose and eliminate the bands corresponding to non-zero values of $k_{z}$. 

The bulk unfolding scheme unfolds the supercell band structure onto the bulk primitive cell, as shown in Fig. \ref{fig:convert}d. We note that the band structure path for bulk unfolding is $X-\Gamma-X$, which is mapped to $\bar{M}-\bar{\Gamma}-\bar{M}$ in the (001) projection, as illustrated in \ref{fig:convert}e.
Bulk unfolding is not straightforward because the matrix $\textbf{M}$ in equations (1) and (2) must be an integer matrix. However, the transformation between a "standard slab" , and the primitive bulk basis vectors may produce a non-integer matrix. The "standard slab" means that the slab's unit cell structure is standardized according to the crystal symmetry and lengths of basis vectors \cite{setyawan2010high}. For instance, the basis vector matrix of a zinc blende primitive cell is given by: 
%as shown in Fig. \ref{fig:matrix}, 
\begin{equation}
    \textbf{a} = \begin{bmatrix}
 0 & x & x\\ 
 x& 0 & x\\ 
 x&x  &0 
\end{bmatrix}
\end{equation}
A (111) slab with one  unit cell in the $xy$ plane and N unit cells in the $z$ direction may be generated by applying the integer transformation matrix \textbf{M}$_{1}$:
\begin{equation}
    \textbf{A}_{1} = \begin{bmatrix}
 0 & -x & x\\ 
 -x& x & 0\\ 
 -2Nx&-2Nx&-2Nx
\end{bmatrix} = \textbf{M}_{1} \cdot \textbf{a}
\end{equation}
Because the basis vectors comprising the matrix \textbf{A}$_{1}$ are not standard, commonly used software packages, such as Python Materials Genomics (pymatgen) \cite{ong2013python} and Atomic Simulation Environment (ASE) \cite{larsen2017atomic}, will automatically apply a transformation, \textbf{M}$_{2}$, into the standard basis vectors:
\begin{equation}
    \textbf{A}_{2} = \begin{bmatrix}
 \frac{x}{\sqrt{2}} &  \frac{\sqrt{3}x}{\sqrt{2}} & 0\\ 
 \frac{x}{\sqrt{2}} &  -\frac{\sqrt{3}x}{\sqrt{2}} & 0\\ 
0 & 0 & 2\sqrt{3}Nx
\end{bmatrix} = \textbf{M}_{2} \cdot \textbf{A}_{1}
\end{equation}
The final transformation matrix, $\textbf{M}_{2}\textbf{M}_{1}$, used to obtain the standard basis vectors comprising the matrix $\textbf{A}_{2}$ is not an integer matrix due to rotation, which prevents the usage of band unfolding. Hence, in order to perform bulk unfolding, the standard slab represented by the matrix $\textbf{A}_{2}$ needs to be converted into the non-standard slab, represented by the matrix $\textbf{A}_{1}$, which can be obtained from the primitive cell via an integer transformation matrix. We have developed an open-source script, compatible with VaspBandUnfolding, that takes the bulk structure and surface Miller indices as input and outputs the correct non-standard slab structure for band unfolding and the integer transformation matrix $\textbf{M}_{1}$. The code is available for download at: \url{https://github.com/Shuyangzero/Unfolding}. 
%spglib \cite{togo2018texttt}  
As seen in Fig. \ref{fig:convert}d, bulk unfolding onto the primitive cell results in a band structure, whose prominent features closely resemble that of the bulk material. The bands corresponding to $k_{z}$ = $|\Gamma X|$ are eliminated, producing a band structure that is directly comparable to ARPES measurements. Surface states arising from the reconstruction appear as weak signatures in the background. The photon energy used in an ARPES experiment determines whether or not such surface states can be detected. From here on, we only compare bulk-unfolded band structures to ARPES experiments. The $xy$-unfolded and $z$-unfolded band structures are provided in the supplemental information.

\subsection{DFT settings}
DFT calculations were performed using the Vienna \textit{ab initio} simulation package (VASP) \cite{PhysRevB.47.558} with the projector augmented wave method (PAW) \cite{PhysRevB.50.17953,PhysRevB.59.1758}. The generalized gradient approximation (GGA) of Perdew, Burke, and Ernzerhof (PBE) was employed for the description of exchange-correlation interactions among electrons \cite{PhysRevLett.77.3865,PhysRevLett.78.1396}. 
A machine-learned Hubbard U correction was applied to the $p$-orbitals of In, As, and Sb. The optimal values of U$_{eff}$ were found by Bayesian optimization \cite{yu2020machine}. For InAs, U$_{eff}^{In, p}$ = -0.5 eV and U$_{eff}^{As, p}$ = -7.5 eV. For InSb, U$_{eff}^{In,p}$ = -0.2 eV, U$_{eff}^{Sb,p}$ = -6.1 eV. With these parameters, PBE+U(BO) yields a band gap of 0.31 eV for InAs and 0.14 eV for InSb, in good agreement with the experimental values of 0.41 eV and 0.24 eV, respectively \cite{Vurgaftman2001}. 
%We note that these U$_{eff}$ values are based on the implementation of the Dudarev formalism in VASP. Different DFT+U implementations may yield different results \cite{Harald_DFT+U}. 
Spin-orbit coupling (SOC) \cite{PhysRevB.93.224425} was included throughout and the energy cutoff was set to 400 eV.  
  
\begin{figure}
\includegraphics[scale = 0.8]{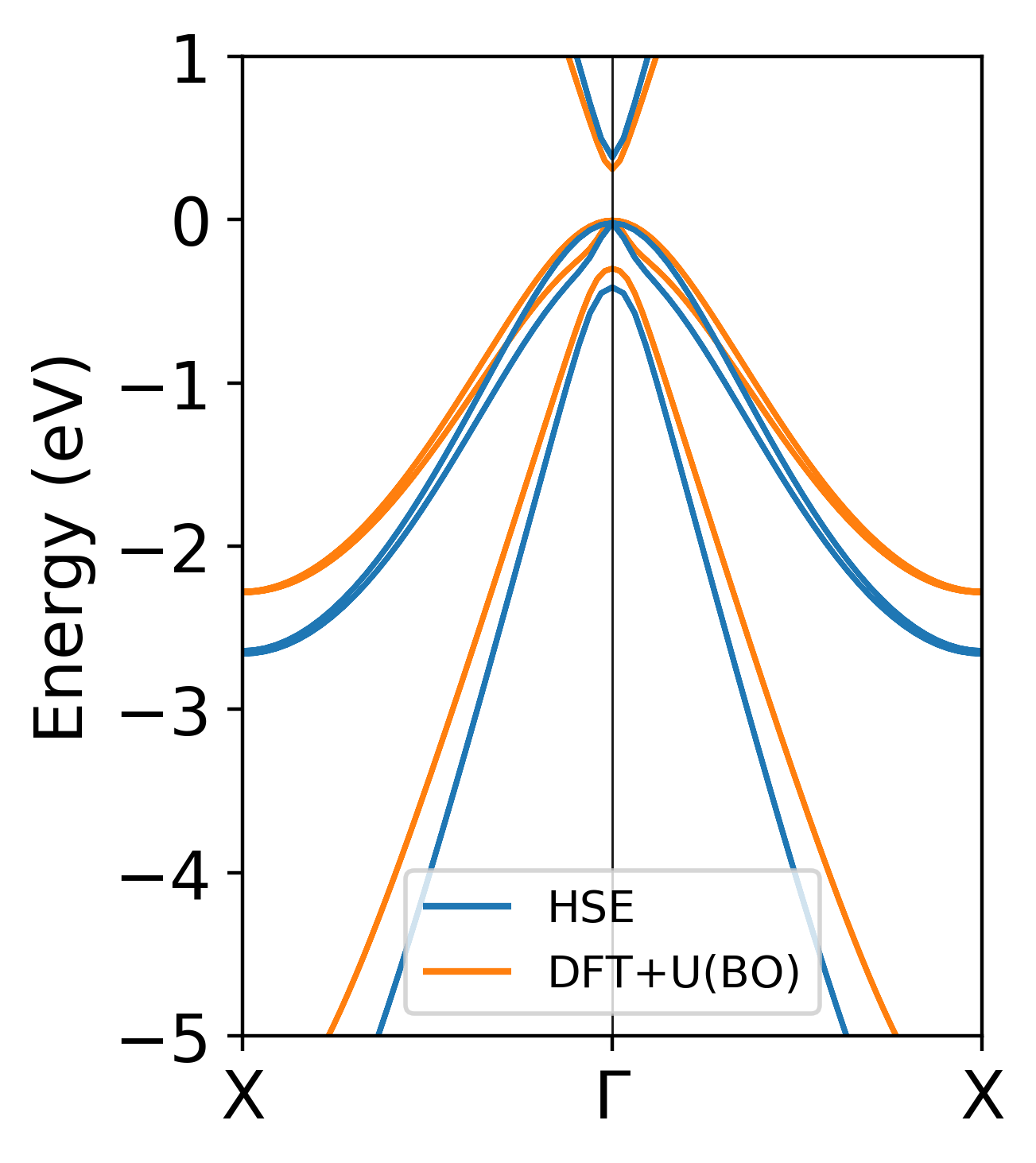}% Here is how to import EPS art
\caption{\label{fig:inas001hse}Band structures of bulk InAs oriented in the (001) direction obtained with PBE+U(BO) and HSE.}
\end{figure}

 For verification, we have compared the band structures obtained with PBE+U(BO) to those obtained with HSE for bulk unit cells of InAs and InSb with the same orientations as the surface slabs studied here. A 8$\times$8$\times$8 k-point grid was used for these calculations. A representative example for InAs(001) is shown in Fig. \ref{fig:inas001hse}. PBE+U(BO) is in close agreement with HSE for the band gap and the band structure features in the vicinity of the Fermi level. However, farther from the Fermi level the agreement deteriorates and the band width is underestimated. The band width is similarly underestimated by PBE+U(BO) away from the Fermi level for all surfaces studied here. 
 
\subsection{Slab construction}
Slab models were constructed using the experimental lattice constants of 6.058 \si{\angstrom} for InAs and 6.479 \si{\angstrom} for InSb \cite{Vurgaftman2001}. 
A vacuum region of about 40 \si{\angstrom} was added in the $z$ direction to avoid spurious interactions between periodic replicas (for the purpose of band unfolding the closest integer number of primitive cells to 40 \si{\angstrom} was used). To avoid surface states due to dangling bonds, the As/Sb atoms on the surface were passivated by pseudo hydrogen atoms with 0.75 fractional electrons and the In atoms on the surface were passivated by pseudo hydrogen atoms with 1.25 fractional electrons. Structural relaxation was performed for the surface In, As, and Sb atoms, the passivating pseudo-hydrogen atoms, and the adsorbed oxygen on InAs(111) until the change of the total energy was below 10$^{-5}$ eV. A 8$\times$8$\times$1 k-point grid was used to sample the Brillouin zone of surface slab models and dipole corrections \cite{PhysRevB.46.16067} were applied. Illustrations of the high-symmetry paths in the Brillouin zone used for surface band structure calculations are provided in the SI.
\begin{figure}[H]
\includegraphics[scale = 0.55]{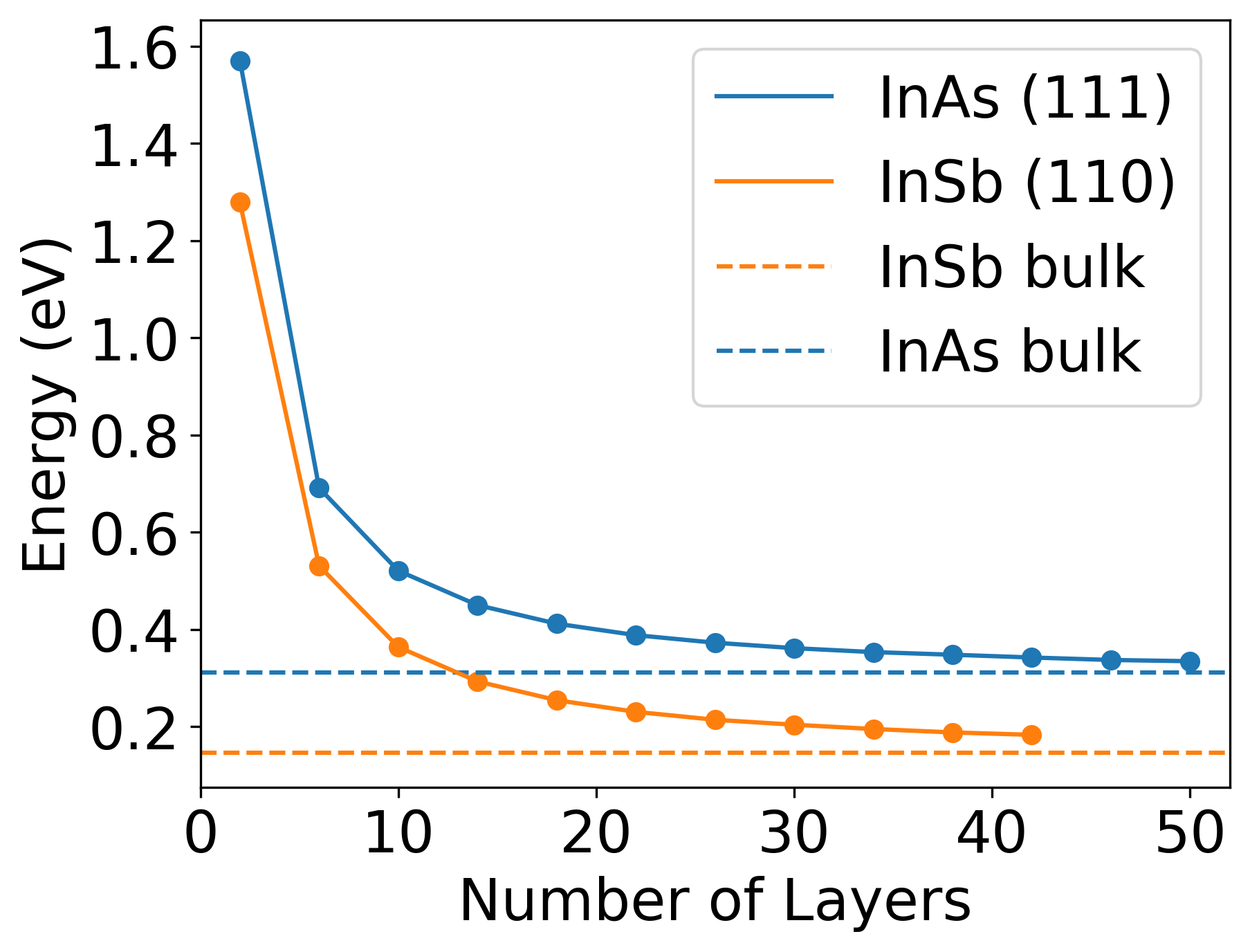}% Here is how to import EPS art
\caption{\label{fig:convergence} The band gap obtained with PBE+U(BO) as a function of the number of layers for InAs(111) and InSb(110) surface slabs.}
\end{figure}
When constructing a surface or interface slab model, it is important to converge the number of layers\cite{PhysRevMaterials.4.034203}. An indicator that the number of layers is converged is that the band gap of the surface slab approaches the bulk limit. Fig.~\ref{fig:convergence} shows the band gap as a function of the number of atomic layers for InAs(111) and InSb(110), as representative surfaces.    
In each iteration the number of layers was increased  by 4. If the band gap difference between the current iteration and the previous iteration was within $5\times10^{-2}$ eV, the current number of layers was regarded as converged. For InAs(111),  50 layers are required to converge the band gap. For InSb(110), 42 layers are required. 
For both surfaces the converged band gap value is close to the bulk value. For InAs(001), modelling surface reconstructions required constructing supercells in the $xy$ plane. Therefore, the maximal number of layers was limited by the computational cost. The largest models we were able to calculate were a 20-layer slab for the  $\beta2(2\times4)$, $\alpha2(2\times4)$ and $\zeta(4\times2)$ surface reconstructions and a 21-layer slab for the $c(4\times4)$ surface reconstruction. These slabs have a reconstructed surface on one side and the other side is passivated with pseudo hydrogen atoms. All of these supercell models contain more than 300 atoms. For the $2\times2$ supercell of oxidized InSb(110), the largest model we were able to calculate is a 22-layer slab. A comparison between the band structures of 42-layer and 22-layer slabs of bare InSb(110) is provided in SI. Although the band gap values are different, no significant difference is found in position of the valence band maximum.

\subsection{Experimental details}

The ARPES measurements were performed at the soft-X-ray ARPES facility \cite{Strocov2014} of the ADRESS beamline \cite{Strocov2010} of the Swiss Light Source, PSI, Switzerland. To maximize the coherent spectral fraction, impaired by the thermal atomic motion \cite{Braun2013}, the measurements were performed at 12K. The combined (beamline and analyzer) energy resolution varied from 50 meV at photon energies h$\upsilon$ around 400 eV to 90 meV around 700 eV, with an angular resolution of the ARPES analyzer of ~0.1$^\circ$. Other relevant details of the soft-X-ray ARPES experiment, including the experimental geometry, can be found in Ref. \cite{Strocov2014}. An intensity transfer function provided in SI has been applied to the ARPES data shown here. The background intensity was removed by subtracting a fraction of the angle-integrated spectrum \cite{Braun2013}. The low energy electron diffraction (LEED) measurements were performed at T=12K with a SPECS ErLEED 150 system in the same vacuum system where the ARPES measurements were performed. 

For InAs(001), the samples were n-type nominally undoped epi-ready wafers supplied by "Wafer Technology Ltd" (UK). The wafers were undoped n-type with a nominal carrier density of n=(1-3)$\times 10^{16}$/$cm^{3}$. They were cleaned in ultra-high vacuum for 20 mins at T=250 $^{\circ} C$ under a stream of atomic hydrogen that was produced by an "MBE-Komponenten" hydrogen atomic beam source (HABS) at a hydrogen pressure of $p\sim 1 \times 10^{-5}$ mbars.

For InSb(110), the samples were n-type nominally undoped epi-ready wafers supplied by "Wafer Technology Ltd" (UK). The samples were cleaved in UHV to expose a fresh, mirror-like surface.

For InAs(111), the samples were n-type nominally undoped epi-ready wafers supplied by "Wafer Technology Ltd" (UK). They were subsequently hydrogen cleaned and an epilayer of InAs was grown in a molecular-beam epitaxy chamber. The samples were then shipped in air to the ARPES facility and cleaned in ultra-high vacuum for 20 mins at T=250 $^{\circ} C$ under a stream of atomic hydrogen at a hyrdogen pressure of $p\sim 1\times 10^{-5}$ mbars.

\section{\label{sec:level1}RESULTS AND DISCUSSION}

\subsection{InAs(001)}

\begin{figure}
\includegraphics[scale = 0.12]{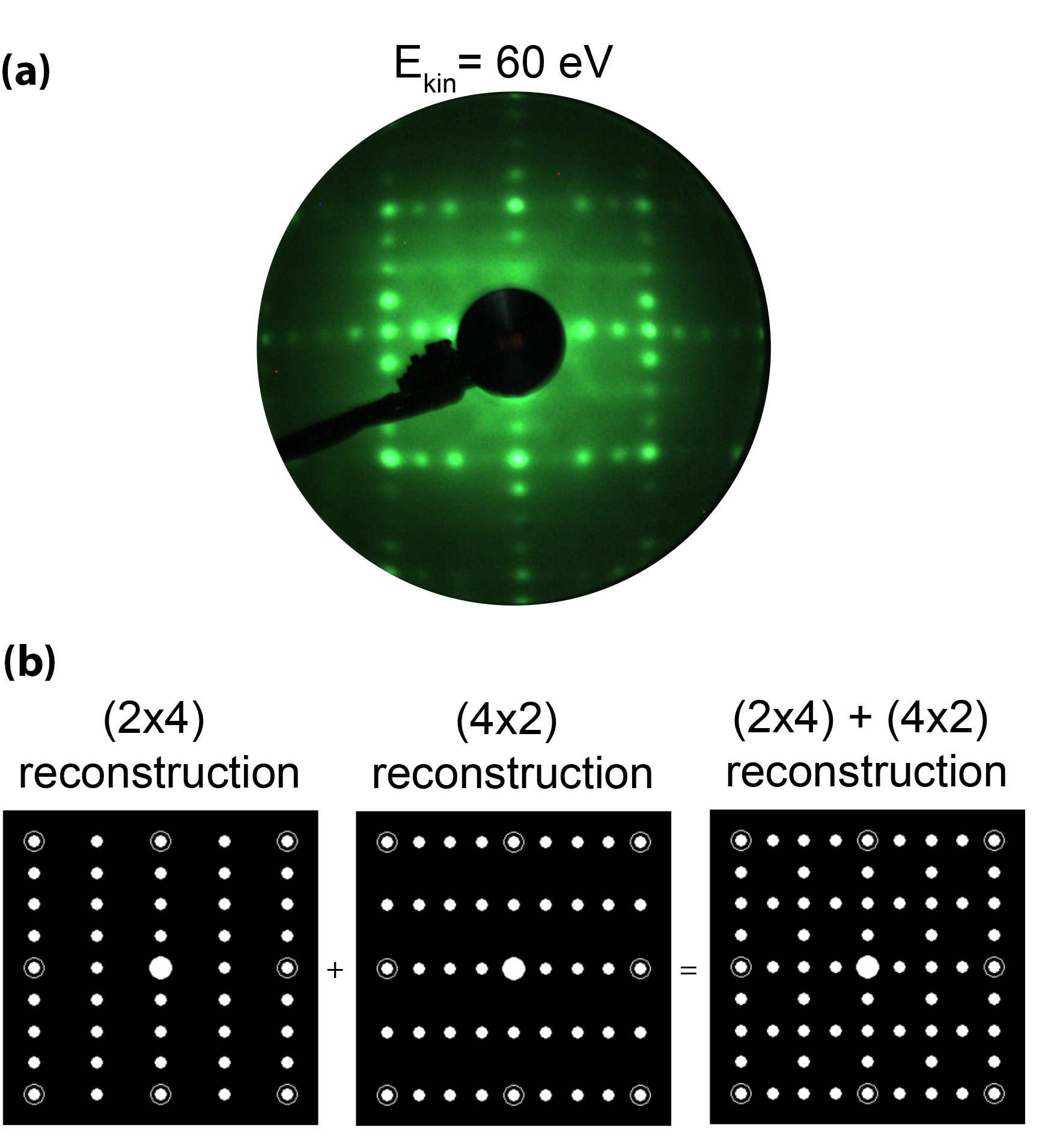}% Here is how to import EPS art
\caption{\label{fig:inas100LEED} (a) LEED results for InAs(001) and (b) simulated LEED patterns corresponding to a 2$\times$4 reconstruction, a 4$\times$2 reconstruction, and a superposition of 2$\times$4 and 4$\times$2 reconstructions.}
\end{figure}
 Fig. \ref{fig:inas100LEED} shows the LEED pattern obtained for the InAs(001) sample. Comparison to simulated patterns suggests coexisting domains of 2$\times$4 and 4$\times$2 reconstructions. This is consistent with earlier reports in the literature that such domains can coexist\cite{bomphrey2017rheed}. The $\times$2 orders appear smeared for both reconstructions, which indicates the presence of disorder at the surface \cite{Laukkanen2012}. Whilst the LEED pattern reveals the size and orientation of the surface reconstruction, the precise atomic structure cannot easily be inferred from it, because LEED is only sensitive to the surface unit cell geometry and not the atomic positions inside the unit cell.

\begin{figure}
\includegraphics[scale = 0.19]{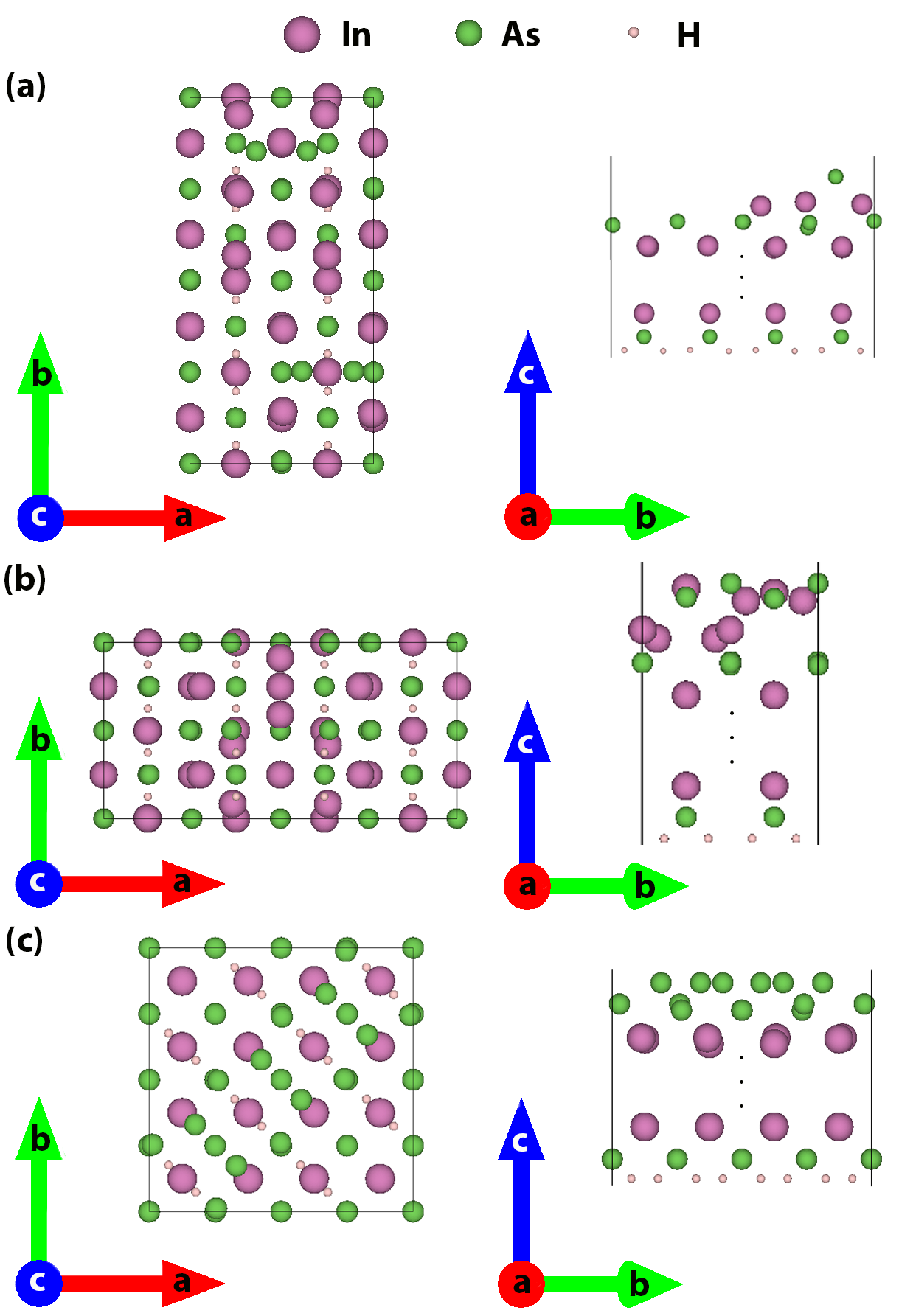}% Here is how to import EPS art
\caption{\label{fig:inas100view} Top view and side view of surface reconstructions of InAs(001): (a) $\alpha2(2\times4)$ (b) $\zeta(4\times2)$ and (c) $c(4\times4)$.}
\end{figure}

\begin{figure}
\includegraphics[scale = 0.12]{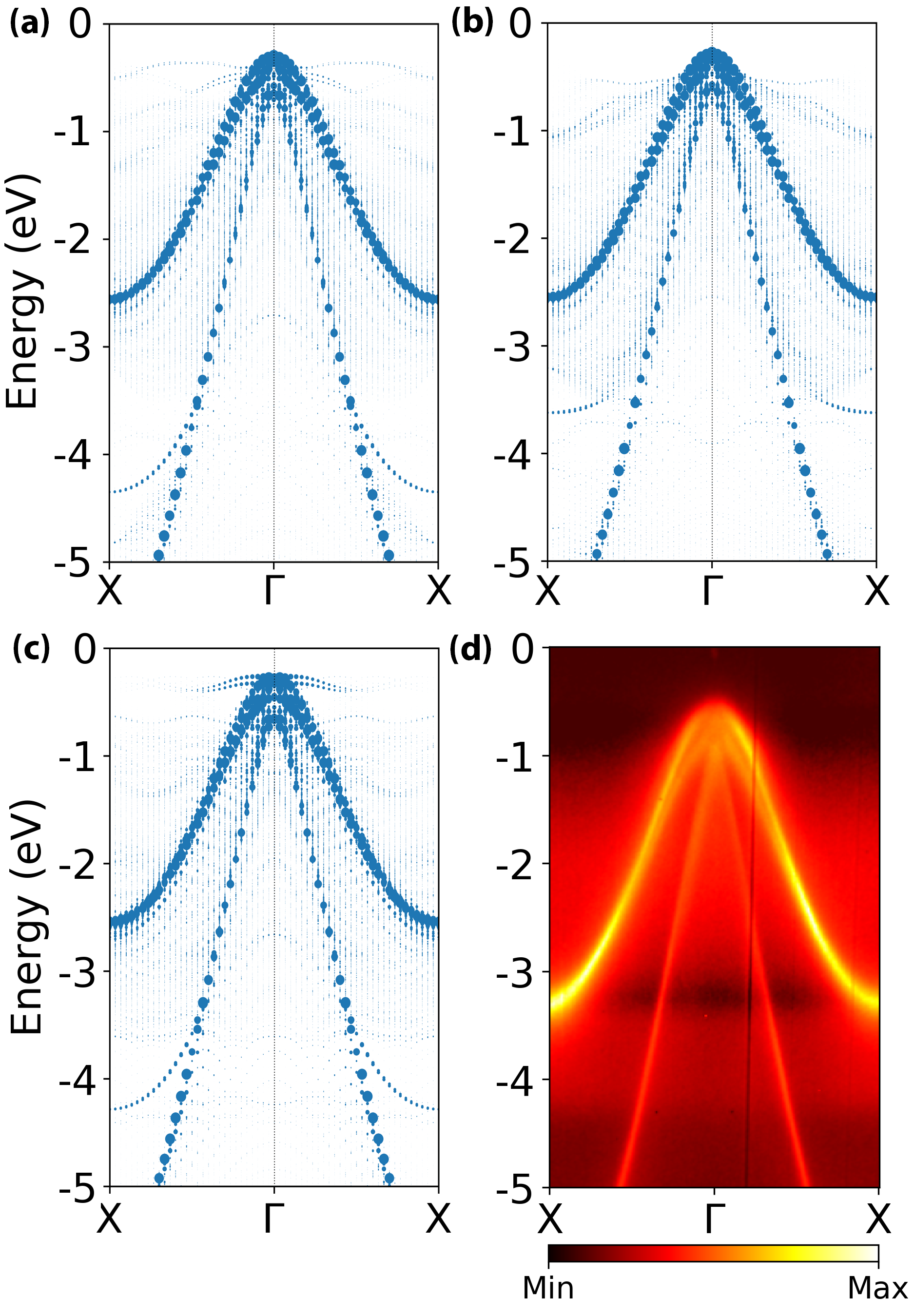}% Here is how to import EPS art
\caption{\label{fig:inas100band} Electronic structure of InAs(001): Bulk-unfolded band structures obtained with PBE+U(BO) for InAs(001) with (a) the $\alpha2(2\times4)$ reconstruction, (b) the $\zeta(4\times2)$ reconstruction, and (c) the $c(4\times4)$ reconstruction, compared to (d) ARPES. The band path for bulk unfolding is $X-\Gamma-X$, which is mapped to $\bar{M}-\bar{\Gamma}-\bar{M}$ in the (001) projection}
\end{figure}

DFT simulations were conducted to investigate the influence of surface reconstruction on the band structure. 
We have focused on the $\beta2(2\times4)$ reconstruction, illustrated in Fig. \ref{fig:convert}a, which is considered as a stable $2\times4$ surface reconstruction in As-rich conditions \cite{yamaguchi1995surface, bell1999species}; the $\alpha2(2\times4)$ reconstruction, shown in Fig.\ref{fig:inas100view}a, which is considered as a stable $2\times4$ surface reconstruction in As-poor conditions \cite{yamaguchi1995surface, bell1999species}; the $\zeta(4\times2)$ reconstruction, shown in Fig.\ref{fig:inas100view}b, which was predicted to be the most stable $4\times2$ surface reconstruction in In-rich conditions by DFT calculations \cite{miwa2003rich}; and the $c(4\times4)$ reconstruction, shown in Fig.\ref{fig:inas100view}c, which was predicted to be stable in As-rich conditions by DFT calculations \cite{ratsch2000surface} and observed experimentally \cite{1014499}. 
The results are compared to the ARPES experiment in Fig. \ref{fig:inas100band}. 
The bulk-unfolded band structures corresponding to different surface reconstructions show distinctive signatures of surface states. The $\beta2(2\times4)$ band structure shown in Fig.\ref{fig:convert}d, the  $\alpha2(2\times4)$ band structure shown in Fig. \ref{fig:inas100band}a and the $\zeta(4\times2)$ band structure, shown in Fig.\ref{fig:inas100band}b, exhibit surface states concentrated mainly around the valley of the heavy hole band between $\Gamma$ and $X$. Both of these reconstructions do not produce surface states near the valence band maximum (VBM).
In contrast, the $c(4\times4)$ band structure, shown in Fig.\ref{fig:inas100band}c, exhibits surface states not only in the valley between $\Gamma$ and $X$ but also at the top of the valence band at the $\Gamma$ point. 

Overall, the main features of the DFT+U(BO) band structures are in good agreement with ARPES. The $\beta2(2\times4)$, $\alpha2(2\times4)$, and $\zeta(4\times2)$ band structures are in somewhat better agreement with experiment than the $c(4\times4)$ band structure because no additional features are experimentally observed near the VBM. We note, however, that the relatively high photon energies used in our experiment (h$\upsilon$=405 eV) are insensitive to the signal from surface states because the extended mean free path of the photoelectrons results in small overlap with states that are sharply localized at the surface \cite{STROCOV2018100}.  Therefore, the comparison of the simulated band structures with different reconstructions to ARPES is inconclusive. It would be possible to better distinguish the type of surface reconstruction from the type of surface states with lower photon energies. In fact, the surface states of InAs(001) $c(8\times2)/2\times4$ have been observed with VUV-ARPES at a lower photon energy of 21 eV \cite{tomaszewska2015surface}.
PBE+U(BO) underestimates the bandwidth of the heavy hole band, consistent with the trend shown in Fig. \ref{fig:inas001hse} and in the SI. Although a hybrid functional would produce a band width in closer agreement with experiment, it is currently unfeasible to conduct such calculations due to the prohibitive computational cost. The top of the VBM in the ARPES is about 0.2 eV lower than in the simulated band structures and a small electron pocket is seen in the ARPES, which is not present in the simulated band structures. This may be caused by the presence of disorder, as indicated by the LEED results, which could lead to Fermi-level pinning \cite{1986_Allen_SurfSci}.

\subsection{InAs(111)}

\begin{figure}
\includegraphics[scale = 0.12]{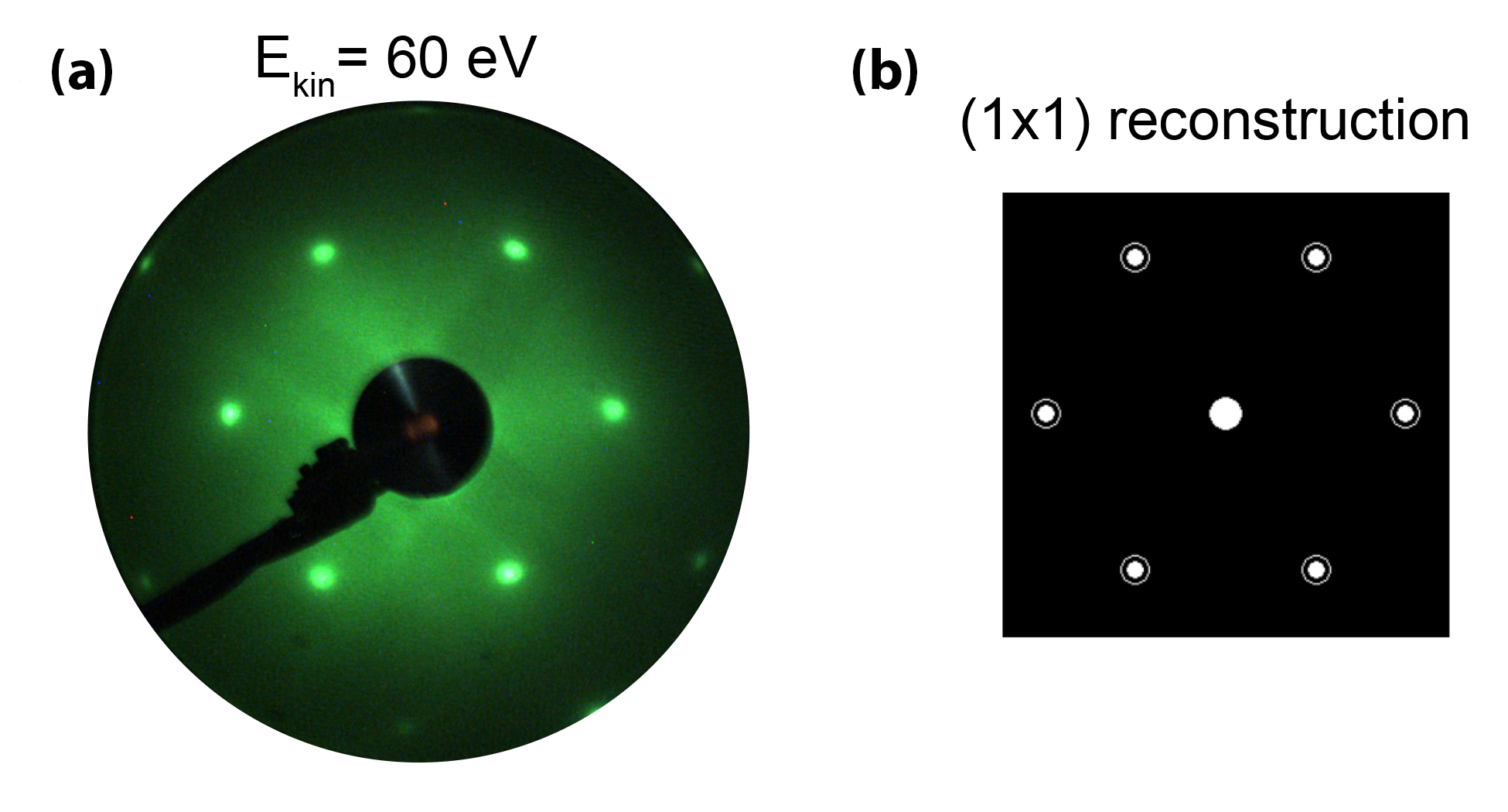}% Here is how to import EPS art
\caption{\label{fig:inas111LEED} (a) LEED results for InAs(111). (b) A simulated LEED pattern corresponding to a $1\times1$ surface}
\end{figure}

\begin{figure}
\includegraphics[scale = 0.12]{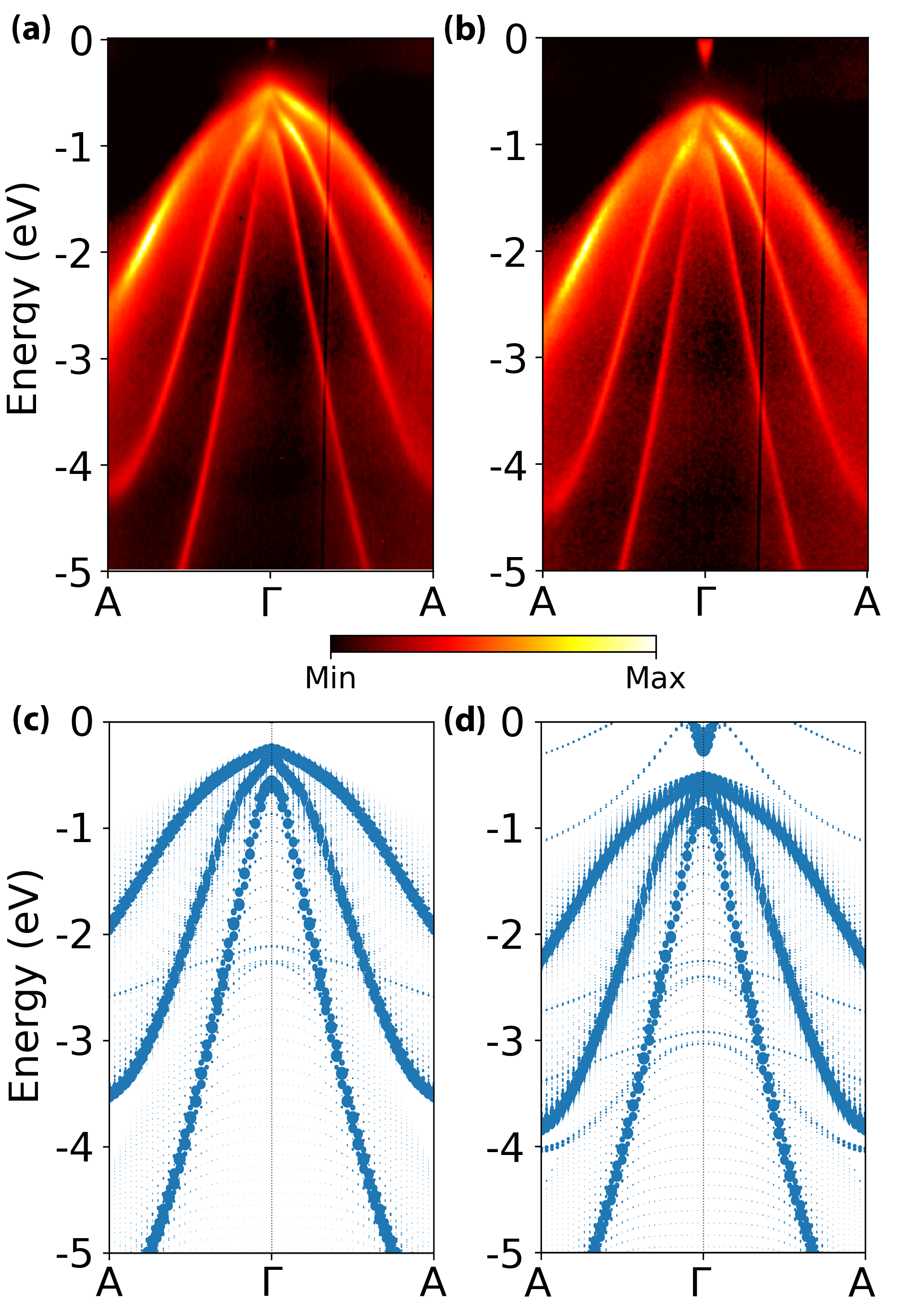}% Here is how to import EPS art
\caption{\label{fig:inas111band} Electronic structure of InAs(111): ARPES results obtained for (a) fresh hydrogen cleaned surface and (b) oxidized surface. The ARPES data is compared with bulk-unfolded PBE+U(BO) band structures of (c) bare InAs(111) and (d) InAs(111)+O. $A$ is the point along $K-\Gamma$ with the coordinates ($\frac{1}{3}$, $\frac{1}{3}$, $\frac{2}{3}$). The band path $A-\Gamma-A$ for bulk unfolding  is mapped to $\bar{K}-\bar{\Gamma}-\bar{K}$ in the (111) projection. The corresponding surface Brillouin zone is provided in SI.}
\end{figure}

  Fig. \ref{fig:inas111LEED} shows the LEED pattern obtained for the InAs(111) sample. Comparison to a simulated pattern indicates that the surface is not reconstructed. Because this sample was grown on the B-side substrate, we assume that the surface is As-terminated. Here we show the results from two ARPES experiments: Fig. \ref{fig:inas111band}a shows a measurement of the bare surface after hydrogen cleaning, whilst Fig. \ref{fig:inas111band}b shows a measurement along the same direction after exposing the surface to ~1800L of pure oxygen in the UHV chamber. One can clearly see that after oxygen exposure, the valence band shifts to higher binding energies and the size of the electron pocket due to charge accumulation increases.
  
\begin{figure}
\includegraphics[scale = 0.19]{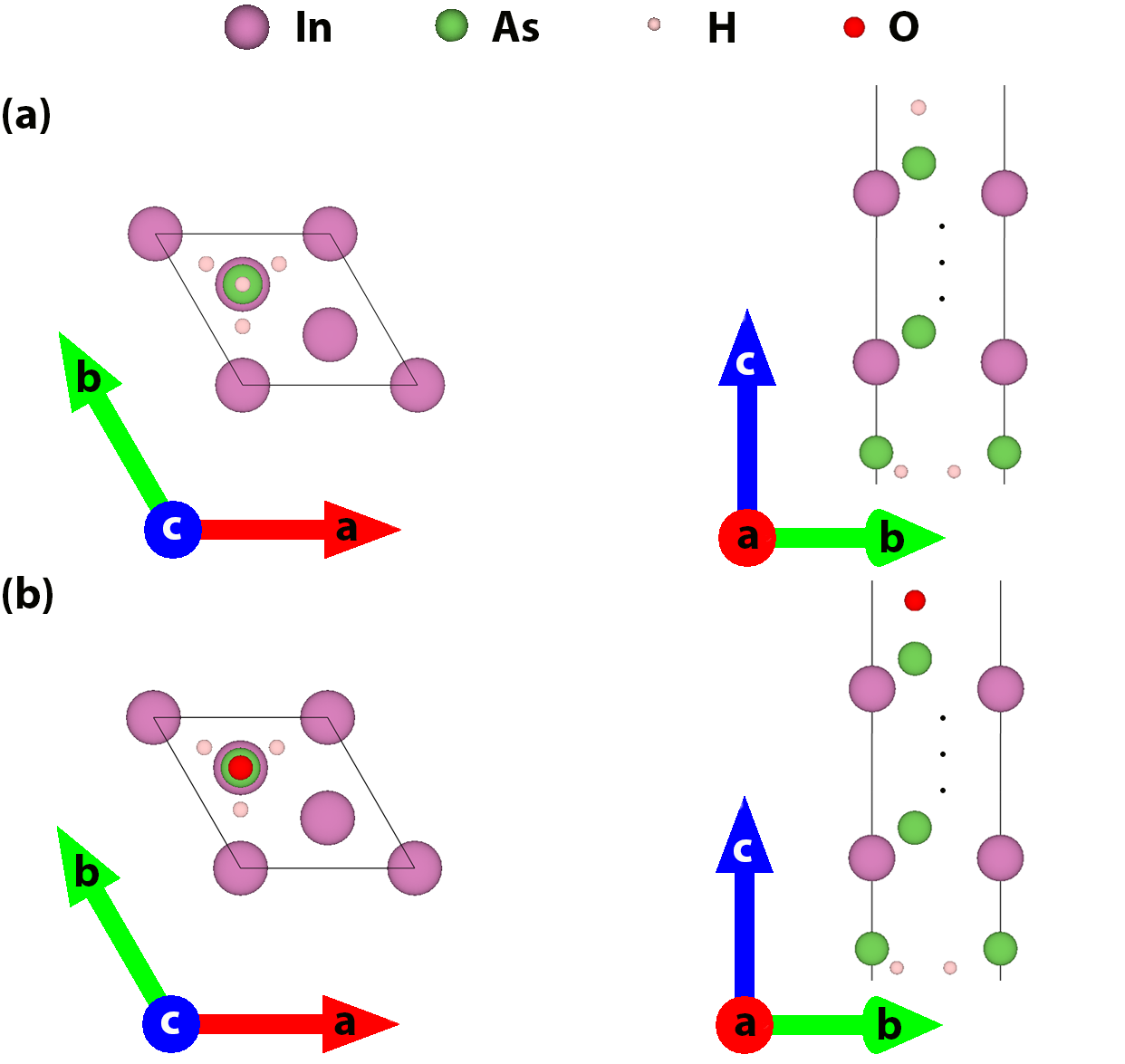}% Here is how to import EPS art
\caption{\label{fig:inas111view} Top and side view of the structure of (a) bare InAs(111) and (b) oxidized InAs(111).}
\end{figure}

To explain the difference between the two ARPES experiments, we constructed two models, as shown in Fig.\ref{fig:inas111view}: a 50-layer As-terminated InAs(111) slab passivated with pseudo hydrogen atoms on both sides and a slab with an oxygen atom bound to As on one side and the other side passivated. We confirmed the bonding between O-As is favorable over the bonding between O-In. The resulting band structures are shown in Fig. \ref{fig:inas111band}c and d, respectively. Overall, the PBE+U(BO) results are in good agreement with ARPES. 
The bandwidth is somewhat underestimated by PBE+U(BO). As shown in \ref{fig:inas001hse} and SI, this could be corrected by using a hybrid functional, however this is currently unfeasible due to the high computational cost. 
 For bare InAs(111) in Fig. \ref{fig:inas111band}c, the VBM is located at -0.24 eV, whereas for oxidized InAs(111) in Fig. \ref{fig:inas111band}d, the VBM is located at -0.47 eV. This indicates that oxidation would lead to band bending, in agreement with the corresponding ARPES results. A significant electron pocket is seen in the simulated band structure of the oxidized surface, in agreement with ARPES. A small electron pocket also is seen in the ARPES data in Fig \ref{fig:inas111band}a, which is not present in the simulated band structure of the bare InAs(111) surface. In this case, there is no evidence of disorder based on the LEED experiment. However, it is possible that the hydrogen cleaning may create some defects that are not taken into account in our simulations.

 To explain why surface oxidation causes band bending, we examined the Bader charge difference between InAs(111)+O and InAs(111). The charge difference for a specific atom A, $C_{diff}(A)$, is given by
\begin{equation}
    C_{diff}(A) = C_{oxidized}(A) - C_{bare}(A)
    \label{eq1}
\end{equation}
where C is the atom's Bader charge. The charge differences of the surface As atom bonded to O and the In atom in the second layer from the surface are -1.08e and +0.11e, respectively. For the oxidized surface, there is significant charge transfer from the surface As atom to the O atom, accompanied by redistribution of the charge on the In atom below the surface. This leads to increased band bending and produces an electron pocket.

\subsection{InSb(110)}

\begin{figure}
\includegraphics[scale = 0.12]{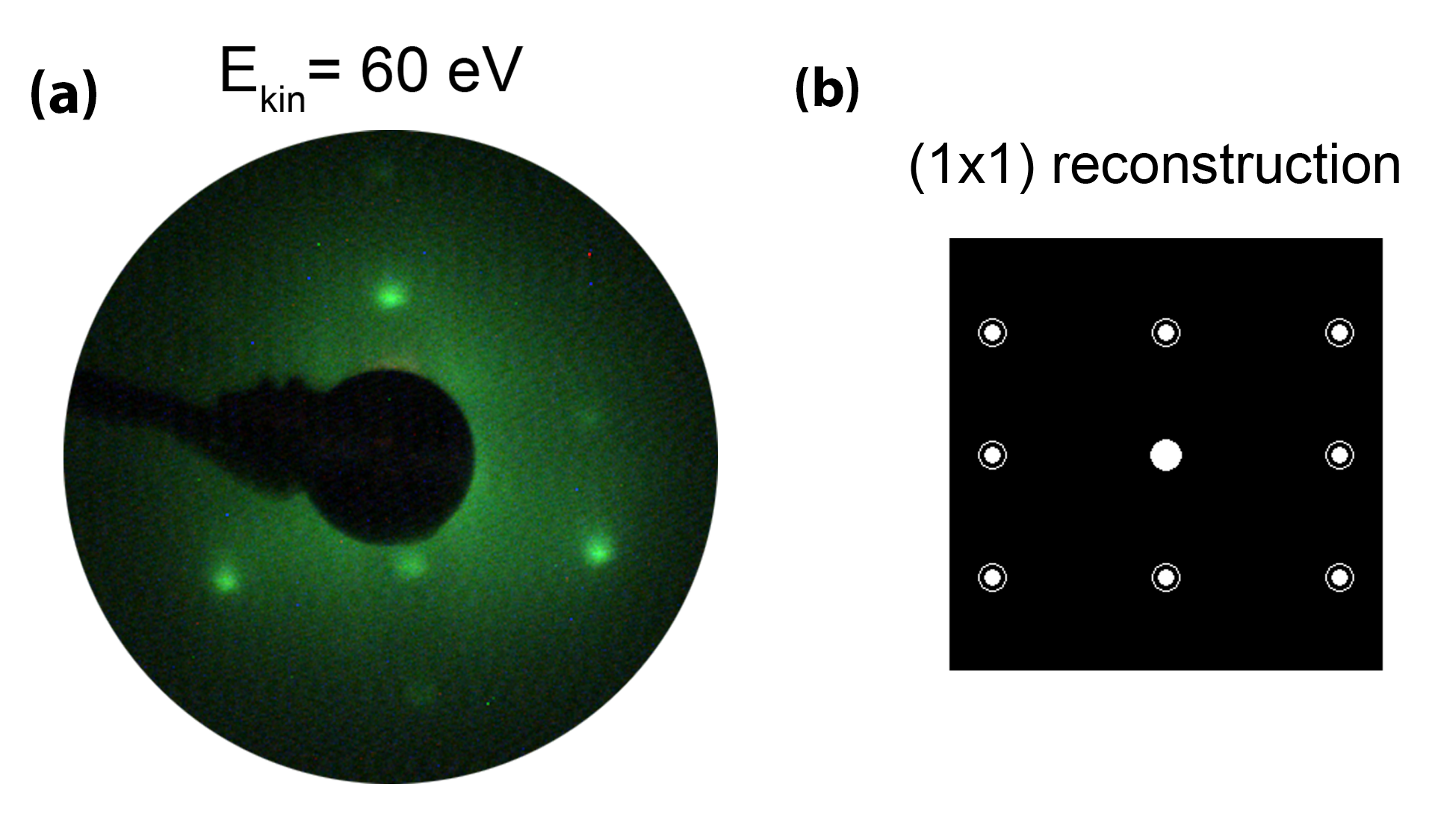}% Here is how to import EPS art
\caption{\label{fig:insb110LEED} (a) LEED results for InSb(110). (b) A simulated LEED pattern corresponding to a $1\times1$ surface.}
\end{figure}

\begin{figure}
\includegraphics[scale = 0.12]{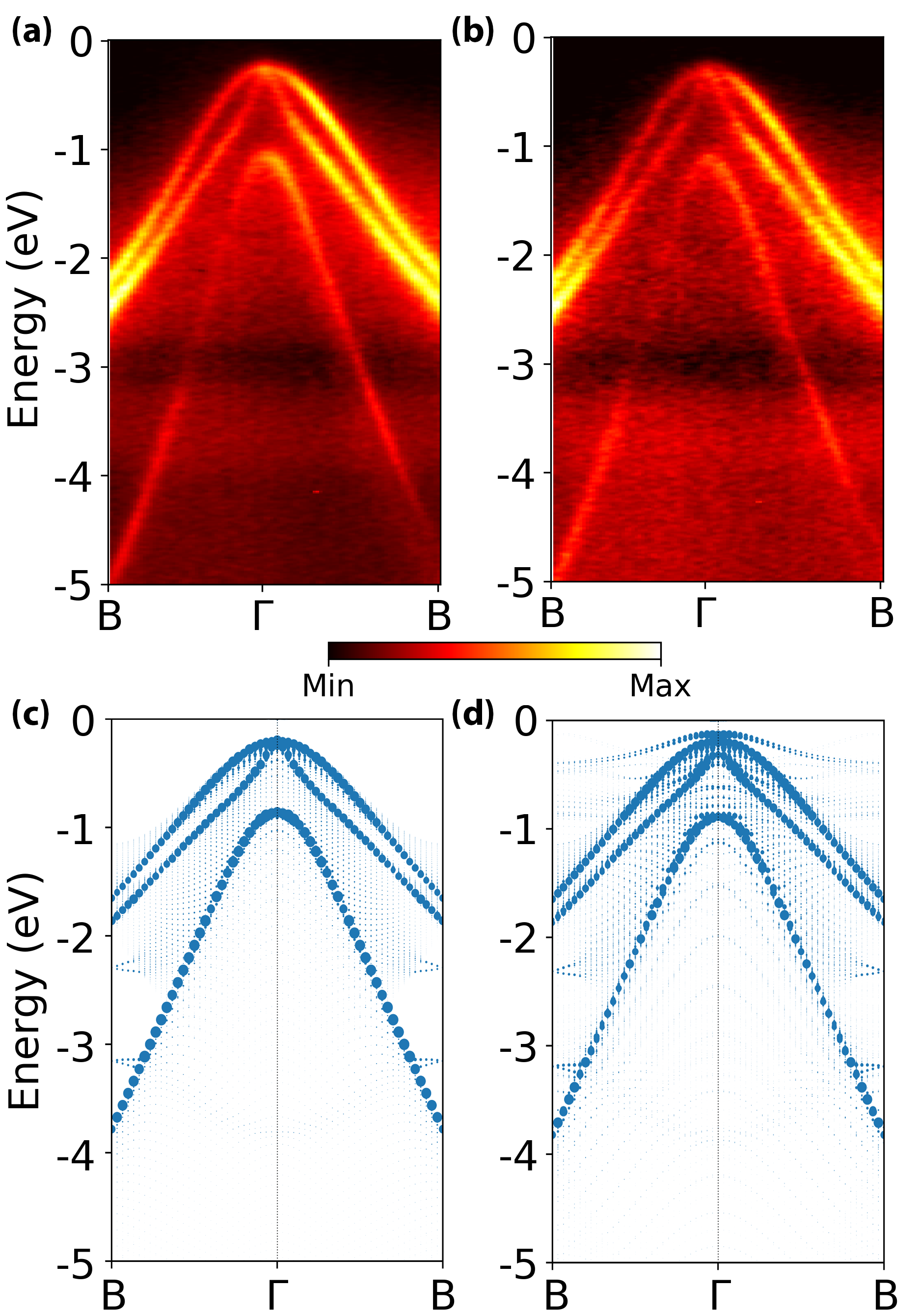}% Here is how to import EPS art
\caption{\label{fig:insbband} Electronic structure of InSb(110): ARPES results obtained for (a) freshly cleaved surface and (b) oxidized cleaved surface. The ARPES  coherent spectral component was enhanced by subtracting the incoherent k-integrated one \cite{Braun2013} approximated by the angle-integrated spectrum,  compared with bulk-unfolded PBE+U(BO) band structures of (c)  bare InSb(110) and (d) InSb(110)+O. $B$ is the point along $X-\Gamma$ with the coordinates (0.25, 0.25, 0). The band path $B-\Gamma-B$ for bulk unfolding  is mapped to $\bar{Y}-\bar{\Gamma}-\bar{Y}$ in the (110) projection. The corresponding surface Brillouin zone is provided in SI}
\end{figure}

Fig. \ref{fig:insb110LEED} shows the LEED pattern obtained for the InSb(110) sample. Comparison to a simulated pattern indicates that the surface is not reconstructed. Similar to InAs(111), here we show the results from two ARPES experiments: Fig. \ref{fig:insbband}(a) shows a measurement of the clean surface after hydrogen cleaning, whilst Fig. \ref{fig:insbband}b shows a measurement along the same direction after exposing the surface to 1350L of pure oxygen in the UHV chamber. In contrast to InAs(111), no significant change of the valence band maximum position and no electron pocket are observed for the oxidized InSb(110) surface.

\begin{figure}
\includegraphics[scale = 0.19]{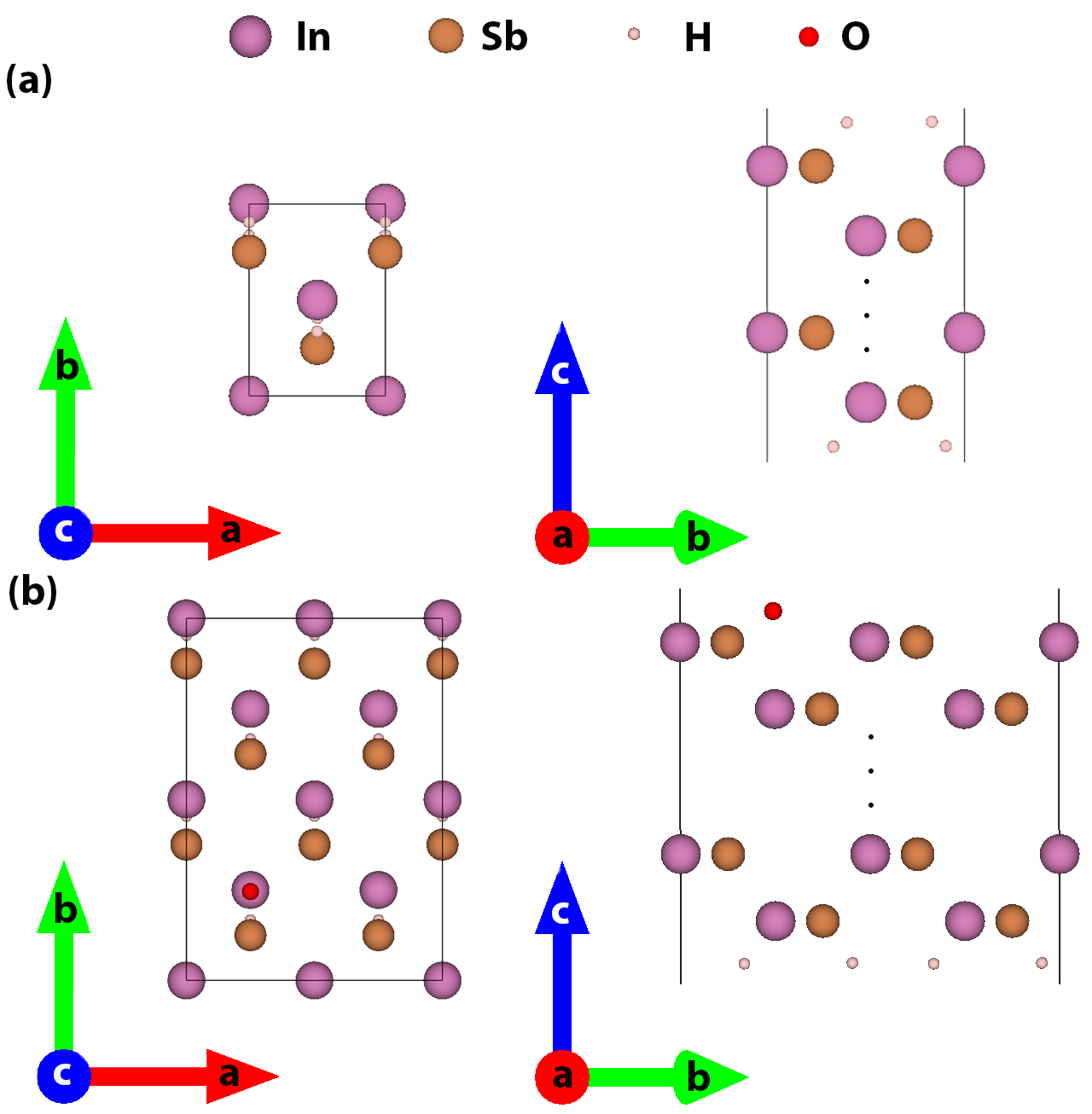}% Here is how to import EPS art
\caption{\label{fig:insbview} Top and side view of the structure of (a) bare InSb(110) and (b) InSb(110)+O ($2\times2$)}
\end{figure}

To simulate this, we constructed two models, as shown in Fig. \ref{fig:insbview}: a 42-layer InSb(110) slab, whose surfaces were passivated with pseudo-hydrogen atoms and a (2x2) supercell slab with an oxygen atom bound to Sb on one side and the other side passivated.  We confirmed that the bonding between O-Sb is favorable over the bonding between O-In. The resulting band structure are shown in Fig. \ref{fig:insbband}c and d, respectively. Overall, the PBE+U(BO) results are in good agreement with ARPES for the main features of the band structure.  Good agreement with ARPES is obtained for the effect of spin-orbit coupling and the position of the split-off band. PBE+U(BO) underestimates the bandwidth and band curvature, consistent with the trend shown in Fig. \ref{fig:inas001hse} and in the SI. Although a hybrid functional would produce a band width in closer agreement with experiment, it is currently unfeasible to conduct such calculations due to the prohibitive computational cost.

  In agreement with ARPES, the VBM position does not differ significantly between the bare InSb(110) and oxidized InSb(110) simulated band structures. In addition, an electron pocket does not appear in the InSb(110)+O band structure, indicating that oxidation does not lead to increased band bending. A surface state that appears in the simulated InSb(110)+O band structure is not observed in ARPES, possibly because of the photon energy used in the experiment is insensitive to surface states. To explain the difference in the effect of oxidation between InAs(111) and InSb(110), we examined the Bader charge difference using equation \ref{eq1}. The charge difference of the surface Sb atom bonded to O is -0.79e and no significant charge difference is found in the second layer below the surface. This indicates that oxidation does not induce as much charge redistribution in InSb(110) as in InAs(111).  Interestingly, the Fermi-level pinning position we find for the oxidized samples is in good qualitative agreement with the branching point theory popularized by Tersoff and others, which predicts Fermi-level pinning in the conduction band for InAs \cite{tersoff1986calculation, 1996_Monch_JApplPhys} and pinning just above the valence band for InSb \cite{tersoff1986failure, 1996_Monch_JApplPhys}.

\section{Conclusion}
In summary, we have studied the electronic structure of the InAs(001), InAs(111), and InSb(110) surfaces using DFT simulations and ARPES experiments. The DFT simulations employed a recently developed method of adding a Hubbard $U$ correction determined by machine learning to a semi-local functional. The optimal values of $U_{eff}$ for InAs and InSb were found by a Bayesian optimization algorithm designed to reproduce as closely as possible the band gap and band structure features obtained with the HSE hybrid functional for the bulk materials. The efficiency of PBE+U(BO) enabled us to perform simulations of surface slab models with several hundred atoms, which would not be feasible with HSE due to its high computational cost. To compare the resulting band structures to ARPES experiments, we have presented a "bulk unfolding" scheme, which unfolds the calculated band structure of a surface slab supercell onto the bulk primitive cell. 

For all three surfaces, the bulk-unfolded band structures calculated with PBE+U(BO) were found to be in good agreement with ARPES experiments. This confirms the transferability of the $U_{eff}$ values obtained for the bulk materials to surface slabs and demonstrates the accuracy of DFT+U(BO). For InAs(001) the effect of surface reconstruction on the electronic structure was studied. The simulated band structures revealed the distinctive signatures of the $\beta2(2\times4)$ reconstruction, the $\alpha2(2\times4)$ reconstruction, the $\zeta(4\times2)$ reconstruction, and the $c(4\times4)$ reconstruction. The $c(4\times4)$ reconstruction is predicted to exhibit a surface state at the top of the valence band, which is not observed in ARPES experiments. Therefore, the results of our simulations support the coexistence of $2\times4$ and $4\times2$ domains. 

For InAs(111) and InSb(110), we studied the effect of surface oxidation on the electronic structure. ARPES experiments and DFT simulations show that oxidation of InAs(111) leads to increased band bending and a larger electron pocket, whereas oxidation of InSb(110) does not significantly affect the position of the valence band maximum and does not produce an electron pocket. This may be attributed to more significant charge transfer from the surface As to O in InAs(111) than from the surface Sb to O in InSb(110). Our systematic theoretical and experimental comparison of surface oxidation in InAs and InSb indicates that charge accumulation is preferred in native oxides of InAs, whilst native oxides of InSb exhibit pinning in the gap, consistent with earlier theoretical results of branching point theory. Our calculations reveal that the microscopic origin of this effect appears to be an increased sub-surface charge redistribution for the native oxide of InAs which is less pronounced in the native oxide of InSb. 

Our results have important implications for proximitized semiconductor nanowire devices. The results elucidate the underlying properties of III-V semiconductor surfaces used to create proximitized p-wave superconductors. In particular, they are important for %understanding The results for bare surfaces shed some light on semiconductor-superconductor interface properties and are important for   
understanding the Fermi-level pinning resulting from reconstructions which, in turn, is relevant for tuning such devices into the topological regime. Similarly, our results on effect of oxidation should underpin thinking about the tunnel junction physics, which controls coupling to Majorana modes. Both of these insights are crucial for the conceptual understanding of device performance as well as for guiding topological qubit design optimization. The results presented here demonstrate that DFT+U(BO) is an efficient and reliable method for simulations of III-V semiconductor surfaces. DFT+U(BO) may be used in the future to conduct simulations of additional surfaces and interfaces. This could advance our understanding of how the electronic structure is affected by atomic structural features.  Therefore, first principles simulations combined with spectroscopy will aid the discovery and realization of new materials systems for quantum computing.

\begin{acknowledgments}
 Work at CMU was funded by the National Science Foundation (NSF) through grant OISE-1743717. This research used resources of the National Energy Research Scientific Computing Center (NERSC), a DOE Office of Science User Facility supported by the Office of Science of the U.S. Department of Energy under contract no. DE-AC02-05CH11231. This work is based on support by the U.S. Department of Energy, Office of Science through the Quantum Science Center (QSC), a National Quantum Information Science Research Center.
\end{acknowledgments}
%\bibliography{ref}% Produces the bibliography via BibTeX.

\end{document}